\advance\topmargin by 2.0in
\documentstyle[aps,prl,epsf]{revtex}
\begin{document}
\twocolumn[\hsize\textwidth\columnwidth\hsize\csname@twocolumnfalse%
\endcsname

\title{Spin ordering quantum transitions of superconductors in a magnetic field}
\author{Eugene Demler$^1$, Subir Sachdev$^{1,2}$, and Ying Zhang$^2$}
\address{$^1$Department of Physics, Harvard University, Cambridge MA 02138\\
$^2$Department of Physics, Yale University, P.O. Box 208120, New
Haven, CT 06520-8120}
\date{\today}

\maketitle

\begin{abstract}
We argue that recent neutron scattering measurements by Lake {\em
et al.\/} (Science {\bf 291}, 1759 (2001)) of the spin excitation spectrum of
La$_{2-\delta}$Sr$_\delta$CuO$_4$ in a magnetic field can be
understood in terms of proximity to a phase with co-existing
superconductivity and spin density wave order. We present a
general theory for such quantum transitions, and argue that their
low energy spin fluctuations are controlled by a singular
correction from the superflow kinetic energy, acting in the region
outside the vortex cores. We propose numerous experimental tests
of our theory.
\end{abstract}
\pacs{PACS numbers:} ]

Recent neutron scattering experiments of Lake {\em et al.}
\cite{gabe} have opened a new window on the spin excitation
spectrum of the high temperature superconductors. They have
observed that a moderate magnetic field induces an anomalously
large increase in the spectral density of low energy spin
fluctuations in La$_{2-\delta}$Sr$_\delta$CuO$_4$ at optimal
doping ($\delta=0.163$) and low temperatures ($T$).  Experiments
on the underdoped La$_{2-\delta}$Sr$_\delta$CuO$_4$ also show a
large increase in the intensity of elastic neutron scattering in
the applied magnetic field \cite{katano,gabe2,boris}. There have
been a number of studies of enhanced antiferromagnetic
correlations in the cores of vortices in the superconducting
order\cite{ss,arovas,hedegard}, and one interpretation of the
measurements is that the extra scattering arises from quasi-static
moments \cite{arovas,hedegard} in the cores of the field-induced
vortices. The measurements then lead to the estimate \cite{gabe}
that the each Cu site in the vortex core has a moment of order
$0.6 \mu_B$ in the anomalous low energy sector. Such a large
moment is characteristic of the insulating antiferromagnet
($\delta=0$), and would require a corresponding charge
disproportionation with a large Coulomb energy cost; this is a
difficulty with such an interpretation.

\begin{figure}
\epsfxsize=2.5in \centerline{\epsffile{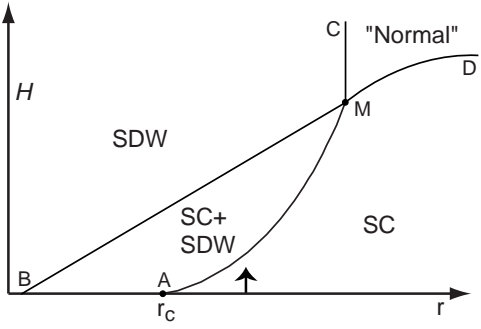}} \caption{Phase
diagram of ${\cal S}+F_{GL}/T$ at $T=0$ and large $N$ as function
of the field $H$ and the parameter $r$ (which is similar, but not
identical to the doping $\delta$) in the limit $\kappa \rightarrow
\infty$. This theory offers a complete, experimentally applicable,
description only of the SC to SC+SDW transition at small $H$, away
from the tetra-critical point M. The remaining phase diagram is
qualitative, and the non-SC phases should have some additional
charge ordering. The path of the experiments of
\protect\cite{gabe} is denoted by the vertical arrow. The upper
critical field of the SC state clearly depends upon $r$, and
$H_{c2}^0$ is its value at $M$. The point A is at the
non-universal value $r=r_c$, but the remaining phase boundaries
can be expressed in terms of $r_c$ and the couplings in ${\cal
S}$\protect\cite{shift}. The point M is at $H/H_{c2}^0=1$,
$r=r_c+v$, the boundary BM is at $H/H_{c2}^0 = 1-(v^2 +
v(r_c-r))/(4u)$, CM is at $r=r_c+v$, and DM is at $H/H_{c2}^0=1+Nv
\Delta/(8 \pi c^2)$, where $\Delta$ is the solution of $\Delta^2 +
(Nu /(2 \pi c^2) \Delta +v -r + r_c=0$. Near M, the position of AM
is given by $H/H_{c2}^0=1-\vartheta_1 (r_c + v -r)/v$ where
$\vartheta_1 = 1.1596 - {\cal O}(v^2)$ ($\vartheta_1 =1.1803 -
{\cal O}(v^2)$) for a triangular (square) Abrikosov flux lattice
(the ${\cal O}(v^2)$ terms will be described elsewhere). The small
$H$ behavior of AM is one of our main results, and is in
(\protect\ref{res2}).} \label{fig1}
\end{figure}

We argue here that the experiments can be understood by assuming
that the superconductor (SC) is in the vicinity of a bulk quantum
phase transition to a state with microscopic {\em co-existence\/}
of SC and spin density wave (SDW) orders; the latter state has
been considered in a number of studies
\cite{inui,so5,bfn,vs,dhl,bala,granath}, and has been observed in
excess-oxygen-doped La$_2$CuO$_{4+y}$ \cite{yslee}. The magnetic
field, $H$, drives the SC phase closer to the SC+SDW phase---see
Fig~\ref{fig1}. Initially, $H$ induces well separated vortices in
the SC, and there could be small additional magnetic scattering
from relatively high energy spin $S=1$ excitons centered around
the vortex cores. However, with increasing $H$ (but with
$H/H_{c2}^0$ still small ($H_{c2}^0$ is the upper critical field
at which the mixed vortex state disappears for a particular set of
parameters--see Fig~\ref{fig1})) the energy of any such exciton
decreases rapidly, and it becomes part of a delocalized collective
spin fluctuation which is a precursor of the transition to the
SC+SDW phase; the dominant magnetic scattering then arises from
the region {\em outside\/} the vortex cores. Our primary results
for extreme type-II superconductors (Ginzburg-Landau parameter
$\kappa \rightarrow \infty$) are: ({\em i\/}) for small
$H/H_{c2}^0$, the SC to SC+SDW phase boundary (MA in
Fig~\ref{fig1}; the caption defines $r$, $r_c$) is at $H \sim
(r-r_c)/\ln(1/(r-r_c))$---so this boundary is anomalously flat at
small $H$, and this allows the system to move close to it for not
too large $H$; ({\em ii\/}) the energy, $\epsilon_{00} (H)$, of
the low energy peak in the neutron scattering cross section should
decrease as $\epsilon_{00} (H) = \epsilon_{00} (0) - C_1
(H/H_{c2}^0) \ln (H_{c2}^0/H)$ along the vertical arrow in
Fig~\ref{fig1}, ({\em iii\/}) when starting from the SC+SDW phase
at $H=0$, one finds an increase of the elastic neutron scattering
given by $I(H) = I(0) + C_2 (H/H_{c2}^0) \ln (H_{c2}^0/H)$, where
$C_1$ and $C_2$ are some constants. All these functional forms are
expected to be exact at small $H$; naively, one might have
expected an analytic series in powers of $\vec{H}^2$, but the
infinite diagmagnetic susceptiblity of a superconductor replaces
this with a non-analytic function of $|\vec{H}|$. We will also
discuss the case of $\kappa$ large but finite.

Because the SC order is present on both sides of the
transition, we can describe it in a {\em static} GL theory: the
free energy $F_{GL}$ is given by
\begin{displaymath}
\int \!\! d^2 x \! \left[\!- |\psi|^2 + \frac{|\psi|^4}{2}  +
\!\left| \left( \vec{\nabla}_x - i \vec{A} \right) \psi \right|^2
\! + \kappa^2 | \vec{\nabla}_x \times \vec{A} |^2 \right].
\end{displaymath}
Here $\psi (x)$ is the complex SC order parameter, and we have
performed standard rescalings to cast this theory in a
dimensionless, universal form: all lengths ($x$) are measured in
units of the (bare) superconducting coherence length $\xi$, the
vector potential is scaled as  $\vec{\nabla}_x \times \vec{A} =
(H/H_{c2}^0) \hat{z}$, and  energies are measured in units of
$H_c^2 \xi^2/(4\pi)$ ($H_c$ is the field at which the free energy
of the uniform superconductor equals that of the normal state).
When $\kappa \rightarrow \infty$, as is the situation in the
experiments on the cuprates \cite{gabe}, the screening of $H$ by
the supercurrents is negligible, and we can develop our theory for
the SC to SC+SDW transition taking $H$ to be uniform. We will not
use any particular model for the dynamics of $\psi$, apart from
assuming that its slow evolution occurs on time scales longer than
those of the spin fluctuations of interest, and is presumably
associated with the thermal and quantum motion of vortices
\cite{fast}.
The SDW order parameter is a $N=3$ component vector,
$\phi_{\alpha} (x, \tau)$, where $\alpha=1\dots N$ and $\tau$ is
imaginary time. The quantum fluctuations of the $\phi_{\alpha}$ as
a function of $\tau$ are known to play an important role even in
the insulating antiferromagnet, and must surely be included in the
vicinity of a quantum transition in a low dimensional magnet. For
simplicity, we will assume  that $\phi_{\alpha}$ is real, but our
theory admits a simple generalization to incommensurate ordering
transitions requiring a complex order parameter \cite{vs}. The
dynamics of $\phi_{\alpha}$ is described by the action
\cite{zeeman} (in the same units as $F_{GL}$):
\begin{eqnarray}
{\cal S} = \int d^2 x \int_0^{1/T} && d \tau \Biggl\{ \frac{1}{2}
\biggl[ (\partial_{\tau} \phi_{\alpha})^2 + c^2 (\nabla_x
\phi_{\alpha})^2 \nonumber \\ && + (r + v |\psi (x) |^2)
\phi_{\alpha}^2 \biggr] + \frac{u}{2} (\phi_{\alpha}^2)^2 \Biggr\}
\label{s}
\end{eqnarray}
where the $x,\tau$ dependence of $\phi_{\alpha}$ is implicit,
$\psi(x)$ is $\tau$ independent as discussed above, and $c$, $r$,
$v$, $u$ are coupling constants, with $v^2 < 4u$. We have
neglected fermionic excitations because momentum conservation
constraints suppress their coupling to $\phi_{\alpha}$ \cite{vs}.
The action is the same as that near SDW ordering
transitions in insulators, with the simplest symmetry-allowed
terms in powers of $\phi_{\alpha}$ and spatial and temporal
gradients.
The tuning parameter $r$ (which, presumably, increases
monotonically with increasing doping, $\delta$) drives the theory
from the SC+SDW phase (with $\langle \phi_{\alpha} \rangle \neq
0$, $\langle \psi \rangle \neq 0$) to the SC phase (with $\langle
\phi_{\alpha} \rangle = 0$, $\langle \psi \rangle \neq 0$) with
increasing $r$; at $H=0$, $T=0$, this transition is at $r=r_c$
(Fig~\ref{fig1}). The latter phase has a $S=1$ exciton (or a
collective ``spin resonance'') associated with oscillations of
$\phi_{\alpha}$ about $\phi_{\alpha}=0$, with an energy which
vanishes as $r \searrow r_c$ \cite{csy}. Our main results for the
$H$ dependence of the physics in the vicinity of the SC to SC+SDW
phase boundary depend crucially on the coupling $v>0$ between the
amplitudes of the SC and SDW orders; such a repulsive coupling was
emphasized in \cite{arovas,so5}.

A powerful tool to account for the quantum fluctuations of
$\phi_{\alpha}$ is the large $N$ expansion \cite{csy}.
At large $N$ and small $T$, the saddle-point equations describing
the properties of ${\cal S} + F_{GL}/T$ in phases with $\langle
\phi_{\alpha} \rangle =0$ are
\begin{eqnarray}
&&{\cal V} (x) = r + v |\psi (x)|^2+ 2N u T \sum_{\omega_n}
G (x,x,\omega_n),\label{e1} \\
&&\left[-1 + |\psi (x)|^2 - (\vec{\nabla}_x - i \vec{A})^2
\right] \psi (x) \nonumber \\
&&~~~~~~~~~+ (NvT/2) \sum_{\omega_n} G (x,x,\omega_n)\psi(x)  = 0
\label{e2}
\end{eqnarray} where $\omega_n$ is a Matsubara frequency
and $G(x,x', \omega_n) \delta_{\alpha\beta} = \int_0^{1/T} d\tau
e^{i \omega_n \tau} \langle \phi_{\alpha} (x,\tau)
\phi_{\beta}(x',0) \rangle$ is the $\phi_{\alpha}$ Green's
function which obeys
\begin{equation}
(\omega_n^2 - c^2 \vec{\nabla}_x^2 + {\cal V} (x)
)G(x,x',\omega_n) = \delta^2 (x-x'). \label{e3}
\end{equation}
The solution of (\ref{e1},\ref{e2},\ref{e3}) (and their
straightforward generalization to phases with $\langle
\phi_{\alpha} \rangle \neq 0 $ \cite{csy}) leads to the phase
diagram in Fig~\ref{fig1}. We emphasize that present model is
accurate only at small $H$ in the vicinity of the SC to SC+SDW
transition. The other phases in Fig~\ref{fig1} involve loss of SC
order, and for these a more complete treatment of the charge
fluctuations is surely needed: various site- and bond-centered
charge ordered states (``stripes'' and ``spin-Peierls'') and
Wigner crystal states are likely to play a significant role (see
{\em e.g.} \cite{vs,granath})

\begin{figure}
\epsfysize=1.75in \centerline{\epsffile{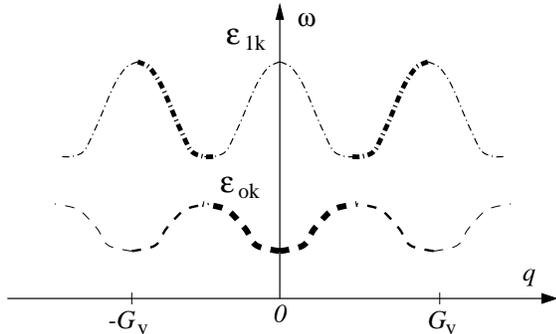}} \caption{
Dynamic spin susceptibility in the vortex state $ \chi''({
q},\omega) = \mbox{Im} [ \int_{x,x'} e^{ i { q} ({x}-{ x}')} G({
x}, { x}',\omega) ]
 = \sum_{n,{ G}_m}
\int_k |\,u_n({ G}_m)|^2\newline\epsilon^{-1}_{nk} \, \delta({
k}+{ G}_m-{ q}) \delta(\omega- \epsilon_{nk}) $, $k$ integration
is over the Brillouin zone of the reciprocal lattice of $\{R_v\}$,
$n$ is a band index, $G_m$ are reciprocal lattice vectors, and
$u_n( G_m)$ are Fourier transforms of the periodic wavefunctions
$u_n(x)$. Note that dispersing $S=1$ exciton modes $\delta(\omega-
\epsilon_{nk})$ are shifted by all $G_m$ with the weight
$|u_n({\bf G}_m)|^2$.} \label{fig3}
\end{figure}

We now present an analytical description of the crucial physical
properties of the solution of (\ref{e1},\ref{e2},\ref{e3}) in the
SC phase at small $H$ and close to $r_c$; a full numerical
solution will be presented in future work. The SC order, $\psi$,
supports an Abrikosov flux lattice of vortices at $N_v \rightarrow
\infty$ positions $\{R_v\}$ ($\psi (R_v)=0$), with a core size of
order unity, and a lattice spacing $L_v \sim \sqrt{H_{c2}^0/H}$.
The resulting $|\psi(x)|^2$ acts like a periodic potential for
$\phi_{\alpha}$, and it is useful express $G$ in terms of the
Bloch states, $g_{nk} (x)=u_n(x) e^{i kx}$ of this periodic
potential:
\begin{equation}
G(x,x',\omega_n) = \frac{1}{N_v} \sum_{nk} \frac{g_{nk}^{\ast} (x)
g_{nk} (x')}{\omega_n^2 + \epsilon_{nk}^2}
\end{equation}
where $(- c^2 \vec{\nabla}_x^2 + {\cal V} (x))g_{nk} (x) =
\epsilon_{nk}^2 g_{nk} (x)$, $n$ is a ``band'' index, $k$ extends
over the first Brillouin zone of the reciprocal lattice of
$\{R_v\}$, and the eigenvalues $\epsilon_{nk}^2  > 0$; SDW order
appears when one of $\epsilon_{nk}$ first vanishes. An immediate
experimental consequence of this structure is the presence of
``Bragg reflections'' at wavevectors separated by the reciprocal
lattice vectors in the dynamic spin structure factor, as shown in
Figure \ref{fig3}.

Our interest will primarily be in the nature of the lowest energy
band, $ \epsilon_{0k}$, which also controls the transition to the
SC+SDW state. As the cores of the vortices act as attractive
potentials for $g_{nk} (x)$, a possibility for such a state is a
superposition of Wannier orbitals localized at the cores of the
vortices \cite{arovas}: $g_{0k} (x) = \sum_{\{R_v\}} e^{ikR_v} f
(x-R_v)$ where $f(x)$ is exponentially localized on a scale $\ell
\ll L_v $. However, the self interactions of the $\phi_{\alpha}$,
accounted for by the self-consistent potential proportional to $u$
in (\ref{e1}), have a large impact on these localized states, and
we will establish that such a structure for $g_{0k}$ must break
down as the transition to the onset of SDW order is approached.
The key argument was made by Bray and Moore \cite{bm} in a
different physical context, and we review their reasoning. As we
are assuming $\ell \ll L_v$, the localized states at the vortex
cores can be treated independently of each other: $\epsilon_{0k}$
is independent of $k$ and $|f(0)| \sim 1/\ell$. Let ${\cal V} (x)
= {\cal V}_>$ (${\cal V} (x) = {\cal V}_<$) for $x$ outside
(inside) the vortex cores where $\psi (x) \approx 1$ ($\psi (x)
\approx 0$). Then subtracting (\ref{e1}) evaluated at these $x$'s
from each other, and noting that the difference in the term
proportional to $u$ arises primarily from the localized state in
the vortex core, we obtain (at $T=0$) $v - Nu |f(0)|^2
/\epsilon_{00} \approx {\cal V}_> - {\cal V}_<
> 0$. This implies that the localization length must be at least
as large as $\ell \sim 1/\sqrt{\epsilon_{00}}$. As $\epsilon_{00}
\searrow 0$ upon approaching the onset of SDW order, we must
eventually have $\ell \sim L_v$, and the assumed structure for
$g_{0k}$ breaks down.

We are, therefore, compelled to turn our attention to extended,
plane-wave like wavefunctions for the $g_{nk}$ (we do not exclude
non-trivial structure in these wavefunctions on the scale $L_v$).
The self interaction terms are now expected to be less important:
we initially neglect the terms proportional to $G$ in
(\ref{e1},\ref{e2}), but will account for them later. We now
demonstrate that, for small $H$, the eigenenergies,
$\epsilon_{0k}$ are controlled by universal structure in $\psi
(x)$ in the region $1 \ll |x| \ll L_v$ well outside the vortex
core at $R_v=0$ (and in the corresponding regions of the other
vortices). Analysis of (\ref{e2}), following the standard
description of a superconducting vortex, shows that in such
regions
\begin{equation}
|\psi (x)| = 1 - 1/(2 x^2); \label{tail}
\end{equation}
the second term above is the correction in the amplitude of the SC
order induced by the superflow kinetic energy. We now use
perturbation theory to evaluate the change in $\epsilon_{00}$
induced by (\ref{tail}); this requires the quantity
\begin{equation}
\langle |\psi (x) |^2 \rangle_x = 1 - (H/(2 H_{c2}^0)) \ln
(\vartheta_2 H_{c2}^0/H), \label{b1}
\end{equation}
where the average is over spatial positions $x$, the logarithm
arises from the integral of $1/x^2$ over the two-dimensional unit
cell of the vortex lattice (it is cutoff at short scales by the
superconducting coherence length $\sim 1$, and at long scales by
$L_v$), and the constant $\vartheta_2 \approx 3$ (for triangular
and square vortex lattices) was obtained by numerical solution of
(\ref{e2}) using the method of~\cite{brandt}. From (\ref{e1}), we
obtain the leading perturbative correction
\begin{equation}
\epsilon_{00}^2 (H) = \epsilon_{00}^2 (0) - (v H/(2 H_{c2}^0)) \ln
(\vartheta_2 H_{c2}^0/H); \label{res1}
\end{equation}
this result is also a variational upper bound on $\epsilon_{00}
(H)$, associated with the wavefunction $g_{00} = 1$. For small
$r-r_c$ we have $\epsilon_{00}^2 (0) = (r-r_c)$ (this corresponds
to the exponent $\nu=1/2$ in mean field theory), and so we can
rewrite (\ref{res1}) as $\epsilon_{00} (H) = (r(H) - r_c)^{1/2}$
where
\begin{equation}
r(H) \equiv r - (v H/(2 H_{c2}^0)) \ln (\vartheta_2 H_{c2}^0/H).
\label{resr}
\end{equation}
The vanishing of $\epsilon_{00} (H)$ determines the position of
the phase boundary AM between the SC and SC+SDW phases at small
$H$:
\begin{equation}
H/H_{c2}^0 = 2 (r - r_c)/[v \ln (1/(r-r_c))] \label{res2};
\label{res3}
\end{equation}
the variational argument shows that this is an upper bound for
$H$, and so the phase boundary can only become flatter at higher
orders. A fully self-consistent calculation which includes the
terms proportional to $G$ in (\ref{e1},\ref{e2}) is more involved,
but tractable: we find a modified functional dependence of
$\epsilon_{00} (H)$ on $r(H)$, $\epsilon_{00} (H) = 2 \pi c^2
(r(H)-r_c) /(Nu (1-v^2 /(4 u)))$ (corresponding to the exponent
$\nu=1$ in the large $N$ limit \cite{csy}), but the expressions
(\ref{resr}) and (\ref{res2}) continue to hold.

We comment further on experimental implications of our results.
For very small $H$, if $g_{0k}$ is localized on a scale $\ell \ll
L_v$, the lowest energy magnetic mode is a relatively high energy
$S=1$ excitonic mode near the vortex cores, as noted earlier. The
spectral weight of this mode will be much smaller than the
slightly higher energy bulk contribution from outside the vortex
cores, and will consist of a feature of width $\delta k =
\ell^{-1}$ (resolution of the Bragg reflections may not be
feasible). For slightly larger $H$, the distinction between the
localized and delocalized contributions will disappear, and we
expect that a full dynamic spin structure with ``Bragg
reflections'' may be resolved in neutron scattering experiments,
with the lowest energy mode obeying (\ref{res1}).  It is also
interesting to note that, even for small $H$, repulsion between
the SC and SDW order parameters described by the $v$ term in
(\ref{s}) will lead to an interesting structure in the spatial
form of $|\psi (x)|$ on the scale $\ell$ which may be
significantly larger than the bare SC coherence length (unity in
our units).  This surprising effect, where the superconducting
vortex core structure depends on the magnetic field for $H \ll
H_{c2}^0$, is another significant prediction of our theory, and is
testable {\em e.g.} in tunnelling experiments.

An interesting recently observed effect \cite{katano,gabe2} may
readily be accounted for by ${\cal S} + F_{GL}/T$: the intensity
of the elastic neutron scattering increases when $H$ is applied to
the SC+SDW state (line BA on Fig~\ref{fig1}). In the large $N$
limit we find that the expectation value of the staggered moment
increases with $H$ as $ \langle | \phi_\alpha | \rangle ^2 = (r_c
- r(H))/(2 u (1-v^2/(4u))) $, with $r(H)$ given in (\ref{resr}).
Quantitative agreement with this relation has been recently
observed\cite{boris}.

\begin{figure}
\epsfysize=1.25in \centerline{\epsffile{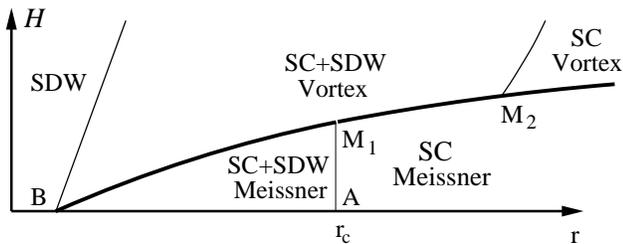}} \caption{Phase
diagram of ${\cal S}+F_{GL}/T$ for finite $\kappa$ and small
fields. The SC and SC+SDW phases in Fig~\protect\ref{fig1} are
identical to the corresponding ``vortex'' phases here. The
position of $H_{c1}$ is given by
$H_{c1}/H_{c2}^0=\sqrt{\pi/2}\,\,(\,\ln\, \kappa/\kappa^2)
\psi_{H=0}^2$, where for the BM$_1$ line
$\psi_{H=0}^2=1-v(r_c-r)/(4u(1-v^2/(4u)))$, and for the M$_1$M$_2$
line $\psi_{H=0}^2 =1+N v \,\epsilon_{00}(0)/8\pi$ with
$\epsilon_{00}$ discussed after equations (\ref{res1}) and
(\ref{res3}).} \label{fig2}
\end{figure}

We now discuss the consequences of supercurrent screening of the
magnetic field at finite $\kappa$. For finite $\kappa$ and small
$H$ we will have a Meissner phase where no vortices penetrate the
sample. As shown in Fig~\ref{fig2}, we may have Meissner phases
that are purely superconducting or have coexisting magnetism. It
is interesting to note that a finite density of vortices
penetrates a sample at $H_{c1}$, so there is a finite jump in the
SDW order across the lines BM$_1$ and M$_1$M$_2$.

We conclude by noting some broader implications of our results.
({\em i\/}) In the discussion above we assumed that $\!v\!>\!0$,
which is consistent with the experimental situation in
La$_{2-\delta}$Sr$_\delta$CuO$_4$. The case  $\!v\!<\!0$, where
the transition into the SC state leads to an enhancement in the
SDW fluctuations, may be also described following the formalism
above. ({\em ii\/}) Closely related phase diagrams should apply to
other ordering transitions of superconductors in a field,
including charge density wave and ``staggered flux'' orders.

This research was supported by NSF Grant DMR 0098226 (Yale), and
by the Harvard Society of Fellows. We acknowledge useful
discussions with G.~Aeppli, R.~Birgeneau, B.~Khaykovich,
M.~Kastner, S.~Kivelson, Y.~Lee, and S.-C.~Zhang.





\begin{thebibliography}{}
\bibitem{fast} There is also a fast mode of $\psi$ oscillations at
frequencies of order the plasma frequency--it is assumed that this
mode merely renormalizes the parameters of the critical modes of
interest here.
\bibitem{zeeman} The spin Zeeman coupling, under which
$\partial_{\tau} \phi_{\alpha} \rightarrow \partial_{\tau}
\phi_{\alpha} - i \epsilon_{\alpha z \beta} H \phi_{\beta}$, leads
to weaker effects of order $H^2$ which we neglect.
\bibitem{shift} We also shift the coefficient of $|\psi|^2$ in
$F_{GL}$ by $(Nv/2) \int d^2 k d \omega /(8 \pi^3 (\omega^2 + c^2
k^2))$ to renormalize away the zero-point fluctuations in
$\phi_{\alpha}$ at $r=r_c$, $T=0$, $H=0$.
\end{thebibliography}
\end{document}